\documentclass[conference]{IEEEtran}
\IEEEoverridecommandlockouts
\usepackage{cite}
\usepackage{amsmath,amssymb,amsfonts}
\usepackage{algorithmic}
\usepackage{graphicx}
\usepackage{textcomp}
\usepackage{xcolor}
\usepackage{svg}

\def\BibTeX{{\rm B\kern-.05em{\sc i\kern-.025em b}\kern-.08em
    T\kern-.1667em\lower.7ex\hbox{E}\kern-.125emX}}
\begin{document}

\title{Non-Invasive Glucose Level Monitoring from PPG using a Hybrid CNN-GRU Deep Learning Network\\
}

\author{
\IEEEauthorblockN{Abdelrhman Y. Soliman}
\IEEEauthorblockA{\textit{Electrical Engineering Department} \\
\textit{Faculty of Engineering,Aswan University}\\
Aswan, Egypt\\
abdelrahman.yahia@eng.aswu.edu.eg}
\and
\IEEEauthorblockN{Ahmed M. Nor}
\IEEEauthorblockA{\textit{Telecommunication Department} \\
\textit{National University of Science and}\\
\textit{Technology POLITEHNICA Bucharest} \\
Bucharest, Romania \\
ahmed.nor@upb.ro}
\and

\IEEEauthorblockN{Octavian Fratu}
\IEEEauthorblockA{\textit{Telecommunication Department} \\
\textit{National University of Science and}\\
\textit{Technology POLITEHNICA Bucharest} \\
Bucharest, Romania \\
octavian.fratu@upb.ro}
\and
\IEEEauthorblockN{Simona Halunga}
\IEEEauthorblockA{\textit{Telecommunication Department} \\
\textit{National University of Science and}\\
\textit{Technology POLITEHNICA Bucharest} \\
Bucharest, Romania \\
simona.halunga@upb.ro}
\and
\IEEEauthorblockN{Osama A. Omer}
\IEEEauthorblockA{\textit{Electrical Engineering Department} \\
\textit{Faculty of Engineering,Aswan University}\\
Aswan, Egypt\\
omer.osama@aswu.edu.eg}
\and

\IEEEauthorblockN{Ahmed S. Mubark}

\IEEEauthorblockA{\textit{Electrical Engineering Department} \\
\textit{Faculty of Engineering,Aswan University}\\
Aswan, Egypt\\
ahmed.soliman@aswu.edu.eg}
}

% btw, the authors not in the same column, a spacing exists 
\maketitle

%\begin{abstract}
%Every year, humanity loses about 1.5 million persons due to diabetic disease. The continuous monitoring of diabetes is massively needed, but the conventional approach, i.e., fingertip pricking, causes mental and physical pain to the patient. Hence, employing a non-invasive blood glucose level monitoring approach is promising for future health care, as it is unpainful and cheaper than the conventional invasive approach. Thanks to the advancement in deep learning, in this papr, we develop a hybrid convolutional neural network-gate recurrent unit (CNN-GRU) network aiming to propose an efficient blood glucose level monitoring mechanism. The CNN is used for extracting spatial patterns in the photoplethysmogram (PPG) signal, while GRU is utilized for extracting the temporal patterns. The proposed model scored MAE of 2.96 mg/dL and RMSE of 3.94 mg/dl and MAPE of 2.40\% and R\textsuperscript{2} score of 0.97.
%\end{abstract}
\begin{abstract}
 Every year, humanity loses about 1.5 million persons due to diabetic disease. Therefore continuous monitoring of diabetes is highly needed, but the conventional approach, i.e., fingertip pricking, causes mental and physical pain to the patient. This work introduces painless and cheaper non-invasive blood glucose level monitoring, Exploiting the advancement and huge progress in deep learning to develop a hybrid convolution neural network (CNN) - gate recurrent unit (GRU) network to hit the targeted system, The proposed system deploys CNN for extracting spatial patterns in the photoplethysmogram (PPG) signal and GRU is used for detecting the temporal patterns. The performance of the proposed system achieves a Mean Absolute Error (MAE) of 2.96 mg/dL, a mean square error (MSE) of 15.53 mg/dL, a root mean square Error (RMSE) of 3.94 mg/dL, and a coefficient of determination ($R^2$ score) of 0.97 on the test dataset. According to the Clarke Error Grid analysis, 100\% of points fall within the clinically acceptable zone (Class A).
\end{abstract}

\section{Introduction}
%Please see the comments as well after %
% Don't do subsections in the introduction if not necessary.
% More coherence is required.
% Write as if it is a story, check the extended version.

%ref before the dot.
%I will read this part when you edit based on the above

Diabetes is a significant global health issue, leading to approximately 1.5 million deaths annually according to the World Health Organization. This places diabetes among the leading causes of death worldwide\cite{WHO}. To prevent diabetes, it is essential to understand its reasons. One major cause is uncontrolled blood glucose level, making its continuous monitoring a crucial task. Traditional methods for monitoring blood sugar, such as micro-needle implantation and finger pricking, can cause both physical and emotional discomfort.

Photoplethysmography (PPG) Signal offers a non-invasive and cost-effective method for obtaining cardiovascular information. The PPG signal is derived from the optical detection of blood volume changes within the skin's microcirculation. LEDs illuminate the tissue, and various biological components deferentially absorb or reflect the light. Fluctuations in blood volume associated with the cardiac cycle cause variations in the intensity of transmitted or reflected light, which are captured as the PPG signal\cite{ppg}. Direct measurement of blood glucose level from the PPG signal is not possible, however, variations in blood glucose can influence factors reflected in the PPG waveform.

% can we abbreviate blood glucose level to BGL?? if applicaple in literature 
% Can you clarify explaination in this paragraph .. make it simple for better understanding
First, \textbf{Vascular Tone Modulation:} Blood glucose level variations can modulate vascular tone, representing the constrictive or relaxed state of blood vessels. These changes can impact blood flow dynamics, as captured by the PPG signal. Secondly, \textbf{Microcirculatory Blood Volume:} Blood glucose level variations can influence blood volume within the microcirculation, the network of minute blood vessels. Elevated glucose concentrations can increase blood viscosity, causing fluctuations in blood flow through these small vessels, which can be reflected in the PPG signal.
% the move from the factor to the proposal doesn't exist, the paper should be a story in harmony, you started by problem statment, importance and motivation, then introduction to pave the way for the proposal (previous 2), then drop?

Table \ref{tab:Related Work} shows different methodologies for non-invasive glucose level monitoring using PPG signals, highlighting the input features and the employed  techniques. As represented in the table previous studies \cite{Fu-Liang} and \cite{Kim} monitored BGL exploiting mathematical Formulas for the features extraction from the frequency domain and time domain, such as Kaiser-Teager energy, spectral entropy, and pulse morphology. The main weak point of \cite{Fu-Liang} and \cite{Kim} is depending on mathematical Formulas to extract features losing the strength of Deep learning extracting the direct relation between the output (BGL) and the input (PPG signals).

This work and LRCN\cite{LRCN} simplify the input features by using only the  PPG signal to build a direct relation between BGL and PPG signals both of Proposed work and LRCN  exploited Convolution Neural Network (CNN) for feature extraction and the ability of recurrent neural network (RNN) to capture temporal dependencies, potentially offering a more efficient and robust solution for non-invasive glucose monitoring. they are inspired by different models. LRCN was inspired by Visual Geometry Group (VGG) \cite{VGG}   to build its CNN meanwhile  Google’s LeNet Inception structure\cite{Lenet} inspired CNN of the proposed model. LRCN used long short-term memory (LSTM) as RNN on the other hand gated recurrent units (GRU) are used in the proposed system as multi-layer GRU outperformed  multi-layer LSTM in terms of training time and accuracy in some applications \cite{lstm_vs_gru}.

% explain your model, structure, use, contribution, purpose, before results
% the reader should feel the novelity and the contribution of the proposal, plus reading brief description about it in the introduction
This new method offers a promising alternative for continuous blood sugar monitoring without the discomfort associated with traditional methods.

% Must mention, the remaining of the paper .....

% re-check related works, don't mention names, be more clear as you can, very simple sentences, few sentences to make it short
\section{Related Work}
Table \ref{tab:Related Work} shows different methodologies for non-invasive glucose level monitoring using photoplethysmography (PPG) signals, highlighting the input features and the employed techniques. The authors in \cite{Fu-Liang} utilize a number of input features, i.e., Fast Fourier Transform, Kaiser-Teager energy, spectral entropy, and pulse morphological characteristics, in addition to physiological and age-related data, to propose new monitoring method. This approach implements a one-dimensional CNN with both micro and macro training strategies to enhance the accuracy of glucose level predictions from the PPG signals. This approach indicates a comprehensive use of signal processing and feature extraction techniques to inform the CNN model, aiming to leverage both morphological and spectral characteristics of the PPG signal for effective glucose monitoring.
While Kim, K.-D's \cite{Kim} work utilizes input features from Intrinsic Mode Functions (IMFs) and employs a variety of machine learning models, including Random Forest (RF), Extreme Gradient Boosting (XGB), CatBoost (CB), and LightGBM.

The proposed method simplifies the input features by using only the  PPG signal by introducing a hybrid Deep Learning network combining Convolutional Neural Network (CNN) and Gated Recurrent Unit (GRU) architectures. This hybrid model aims to utilizes the CNN's strength in feature extraction and the GRU's ability to capture temporal dependencies, potentially offering a more efficient and robust solution for non-invasive glucose monitoring.
% don't explain anythingg related to your proposal in related work section
% section 3 can be proposed methodology, then first paragrapgh about the comparision to previous works (above one)

\begin{table}[h]
    \centering
    \caption{Comparison of PPG-based Non-Invasive Glucose Level Monitoring with the Proposed work.}
    \label{tab:Related Work}
    \begin{tabular}{ccc}
        \hline
        Author(Year)& Input Features  & Method \\
        \hline
        Fu-Liang Yang (2021) \cite{Fu-Liang}&  Fast Fourier transform, &  One dimensional\\
        &  Kaiser-Teager energy,&CNN with micro \\
        &spectral entropy,&and macro training\\
        & pulse morphological,& \\
        & physiological,age&\\
        \hline
         LRCN(2023)\cite{LRCN}& Only the PPG signal&Hybrid \\
        & & CNN-LSTM\\
        & & Deep Learning\\ 
        & & Network\\
        \hline
        Kim, K.-D (2024)\cite{Kim}&  Extracted from IMFs,& RF, XGB,\\
        &  ratio features from PPG& CB, LightGBM\\
         \hline
        Our Method (2024)& Only the PPG signal& A Hybrid CNN-GRU\\
        && Deep Learning Network\\
    \end{tabular} 
\end{table}

\section{Methodology}
\subsection{Dataset Description}

The public Mazandaran dataset version 2 \cite{dataset} is used which includes 67 raw PPG signals sampled at a frequency of 2175 Hz. Each signal corresponds to a blood glucose level measurement from one of the 23 participants. % lasr sent. is not clear

Blood glucose levels were obtained using a standard invasive method with an Accu-Check™ Active Series self-monitoring device and a 333 calibration kit.

% no need for the following
The PPG signals were acquired using the following setup:
\begin{itemize}
\item A green LED transmitter with a wavelength of 550 nanometers (nm)
\item An APDS9008 photodiode receiver
\item A low-frequency amplifier (MCP6001) to boost the signal
\item An RC filter to manage signal amplification and minimize noise
\item A rectifier diode to convert the signal
\item An Atmega328™ microcontroller for analog-to-digital conversion (ADC)
\end{itemize}

\begin{figure}[htbp]
\centerline{\includegraphics[width=0.4\textwidth]{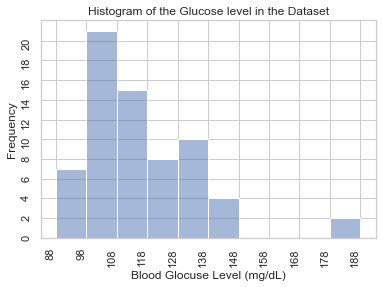}}
\caption{Histogram of the Dataset}
\label{Histogram of the Dataset}
\end{figure}

As illustrated in Figure \ref{Histogram of the Dataset} , the majority of blood glucose level samples fall within the range of 98-138 mg/dL. Meanwhile, 7 samples fall below 98 mg/dL, and 6 samples exceed 138 mg/dL.

% table and histogram explain same concept???? delete one, i prefer deleting fig, or keep it but improve it as i wrote down, plus make it smaller
%\begin{table}
%\centering
%\caption{The Dataset Distribution}
%\label{tab: The Dataset distribution}
%\begin{tabular}{cc}
%\hline
%Range & Frequency \\
%\hline
%88-98 & 7 \\
%98-108 & 21 \\
%108-118 & 15 \\
%118-128 & 8 \\
%128-138 & 10 \\
%138-148 & 4 \\
%178-188 & 2 \\
%\hline
%\end{tabular}
%\end{table}

% big comparable to the included info, not good resolution, colors are not clear.

%% isn't it a story 
\subsection{Signal Preprocessing}
\begin{figure*}[htbp]
\centerline{\includegraphics[width=0.8\textwidth]{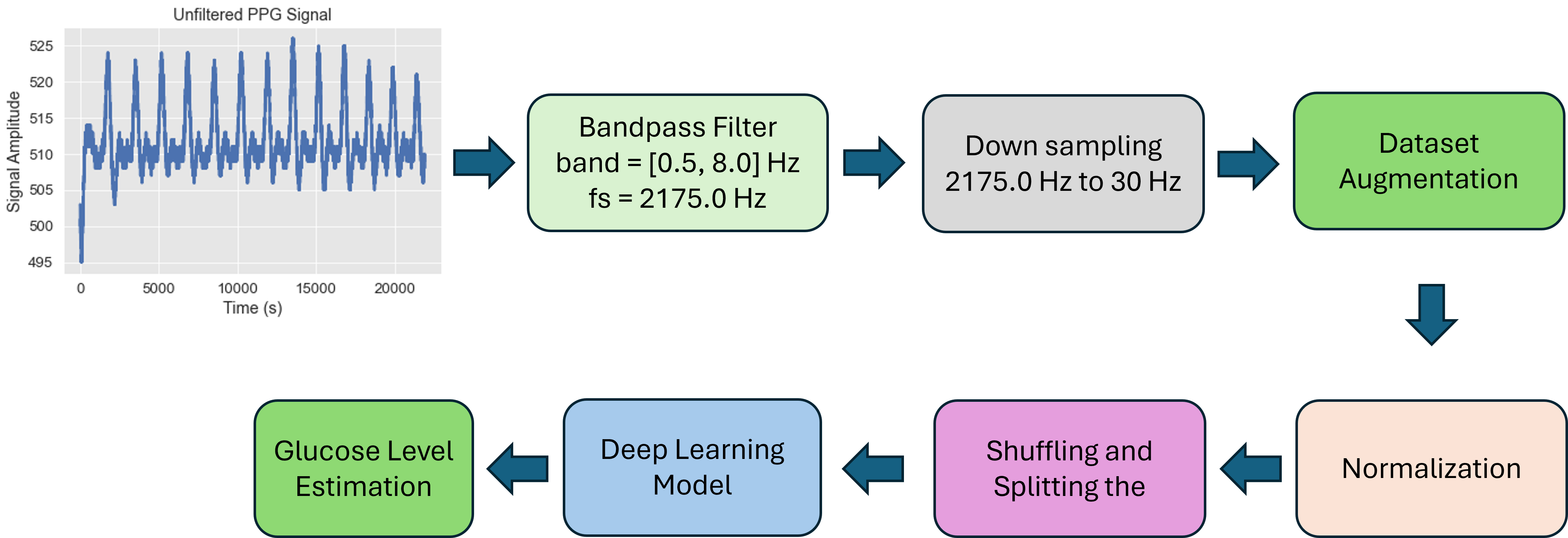}}
\caption{Proposed system architecture.}
\label{Proposed method steps}
\end{figure*}

Figure \ref{Proposed method steps} illustrates the proposed process flow. First, the raw PPG signal goes through pre-processing steps, to prepare the data for the model, which uses the processed signals to estimate glucose levels.
% raw!!, not clear which steps, the following??, which refer to what?
% To prepare the data for the model, the raw PPG signal goes through pre-processing steps, then ???

% blocks font is small, at least 8 times new roman 
% very small fig is not clear, you can make it in two columns 

\subsubsection{Band-pass Filtering}\label{AA}
The unwanted noise is removed from the raw signal using a band-pass filter, which passes the frequencies between 0.5 Hz (30 beats per minute) and 8 Hz (240 beats per minute) as shown in 
\begin{figure}[htbp]
\centerline{\includegraphics[width=0.5\textwidth]{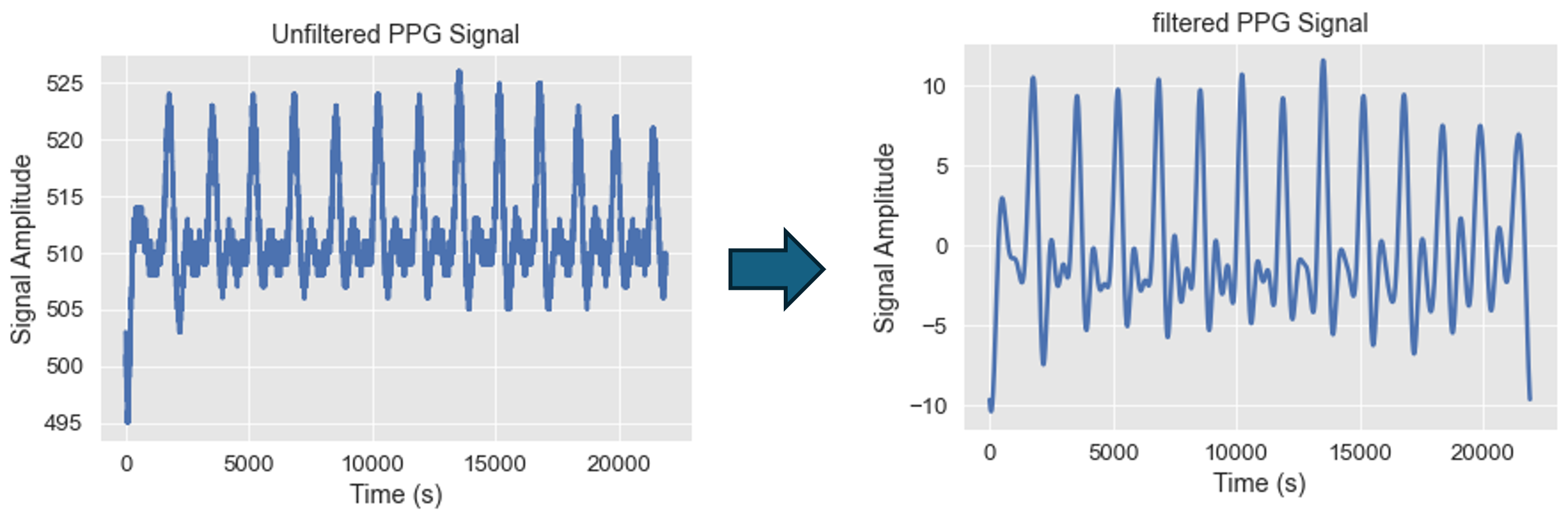}}
\caption{Applying Filter.}
\label{Downsamping}
\end{figure}
% can make it in two rows.
% not clear figures

\subsubsection{Downsampling}
Then, the sampling rate of the signal is reduced from 2175 Hz to 30 Hz aiming to remove unnecessary data, because the highest useful frequency is 8 Hz. According to the Nyquist principle, the sampling rate needs to be at least twice the highest frequency to capture all the information in the signal. Consequently, 30 Hz is a sufficient sampling rate since the filtered signal only contains components below 8 Hz as shown in \ref{Downsamping}.

% fig 3, 4 and 5, can be combined in one figure in one column layout for better sizing benefits, and avoid repetations 
\begin{figure}[htbp]
\centerline{\includegraphics[width=0.5\textwidth]{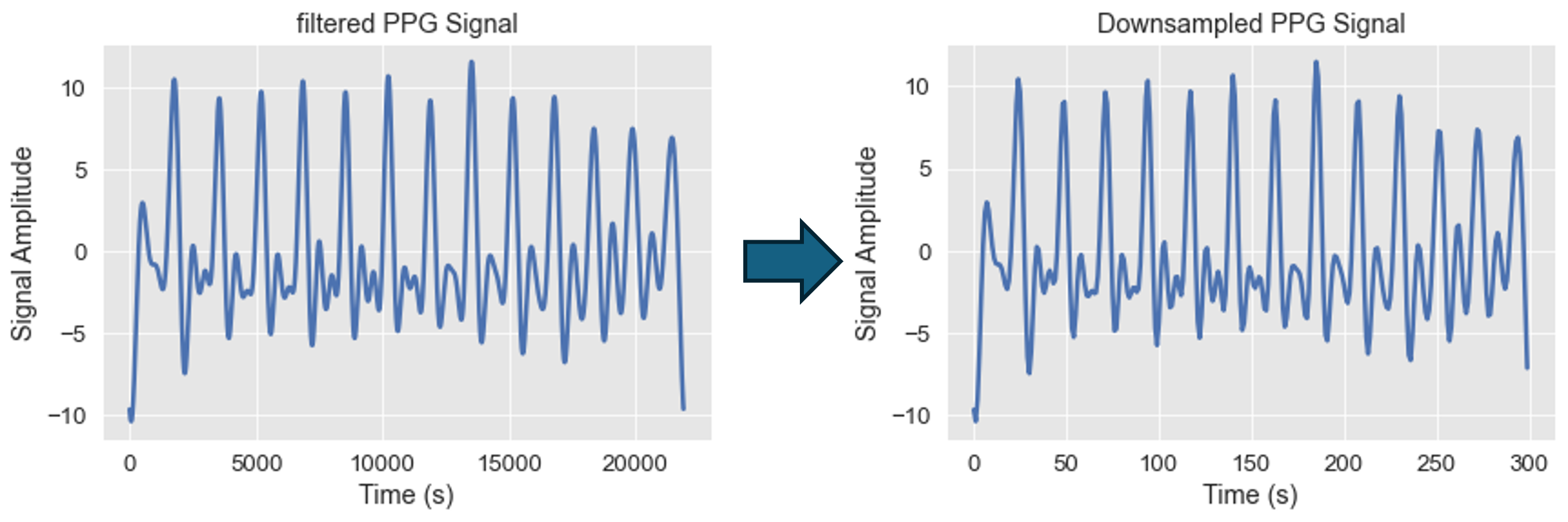}}
\caption{Applying Downsampling.}
\label{Downsamping}
\end{figure}
\subsubsection{Dataset Augmentation}
The sample numbers are increased from 67 to 269 by adding different Gaussian noises to the training PPG signals. Those types of noises follow a normal distribution and were introduced with different standard deviations.
% rewrite this part, not clear at all, though i know how is your method
% remove comment if you have done it :), i don't remember about what i wrote comments anyway 

\subsubsection{Normalization}
\begin{figure}[htbp]
\centerline{\includegraphics[width=0.5\textwidth]{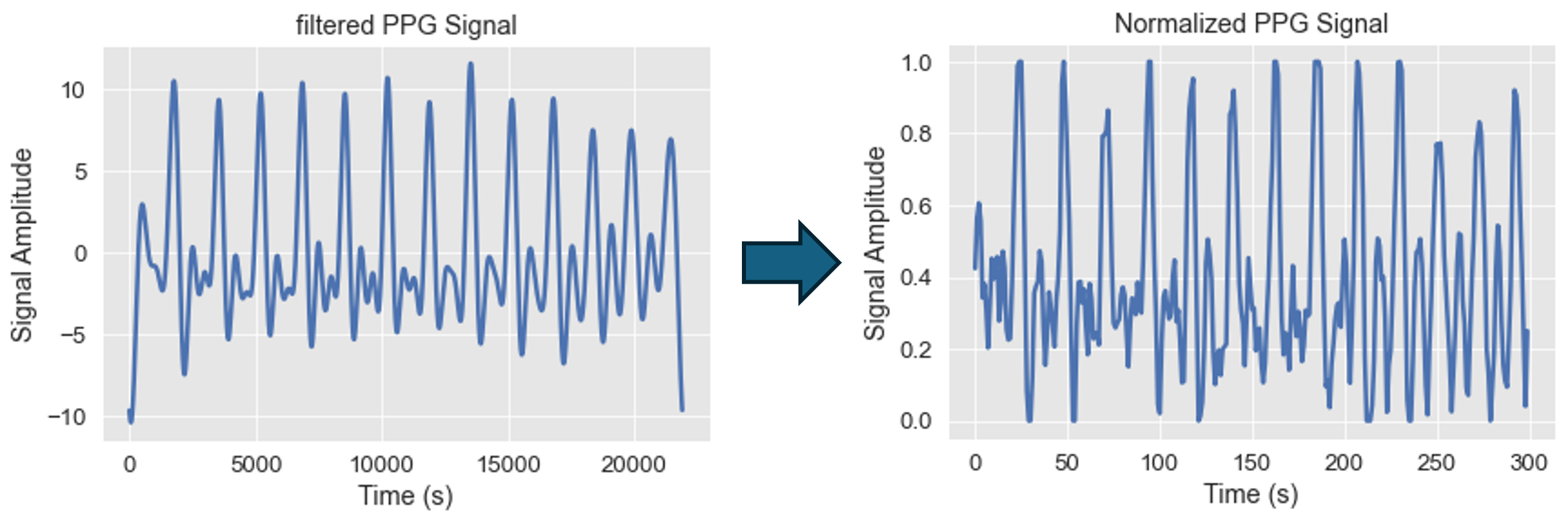}}
\caption{Applying Normalization.}
\label{Normalization}
\end{figure}

Normalization is a crucial preprocessing step in machine learning. Its primary goals are to ensure fair treatment of all features, promote smoother learning by preventing large features from dominating gradient updates, and create a favorable environment for activation functions. By scaling features appropriately, normalization contributes to efficient model training and better convergence as shown in \ref{Normalization}.

% mention which type of scaling, max-min, to unity, mean ??? figure says to unit 1, but not enough 

\subsubsection{Shuffling and Splitting the dataset}
To prevent bias and improve generalization, we shuffle the dataset before training our model. This ensures that the model sees a mix of data points, preventing it from learning based on a specific order. 

Additionally, we split the data into training, validation, and test sets. The model learns from the training set, and the validation set helps us fine-tune the model to avoid overfitting (memorizing the training data too well). 

Finally, the unseen test set provides a final evaluation of the model's ability to perform well on entirely new data.

% I feel more that this should be with the dataset part
% same comment and you didn't solve this
% plus, you explain the proposal, not teaching the purpose of training, validation, and test parts, this is basic, who don't know this sould go to books not a scientific paper

% Is the model weight similar to preprocessing steps??? !!!!
% Make it C or something to clarify it
\subsubsection{Model Architecture}
Inspired by  \cite{Fu-Liang} and Google's LeNet\cite{Lenet} Inception structure,
the proposed architecture leverages the strengths of (CNN) feature extraction and 
Gated Recurrent Units (GRU) in temporal pattern recognition for PPG signals analysis.
% talk more about your network, advantages, dis adv., benefits of cooperation and so on.

The proposed hybrid network as in \ ref{model }incorporates three parallel blocks:

\begin{itemize}
\item \textbf{Two CNN blocks:} These blocks use filters of varying lengths, i.e., kernel sizes, to extract features from the PPG signal. Each CNN block consists of a one-dimensional (1D) convolution layer followed by batch normalization, ReLU activation, and max pooling. 
\item \textbf{One GRU block:} This block uses multiple GRU units with varying numbers of units to identify sequential patterns in the PPG data.
\end{itemize}

Combining these blocks adds the following merits to the network:
\begin{itemize}
\item Each of the three parallel blocks is flattened and then fed into three identical fully connected layers.
\item The outputs from these three branches are then concatenated (merged) before a final fully connected layer.
\item The final layer is a single unit for regression, providing a continuous output value.
\end{itemize}
% I don't feel it is strengths, i mean this is not clear in sentences, it is like describing for steps or something will be done by the network.
% rewrite please

In essence, this architecture combines the feature extraction capabilities of CNNs with the sequential pattern recognition power of GRUs for improved PPG signal analysis. 
% This sentence is a repetition? 

    \begin{figure}[htbp]
    \centerline{\includegraphics[width=0.6\textwidth]{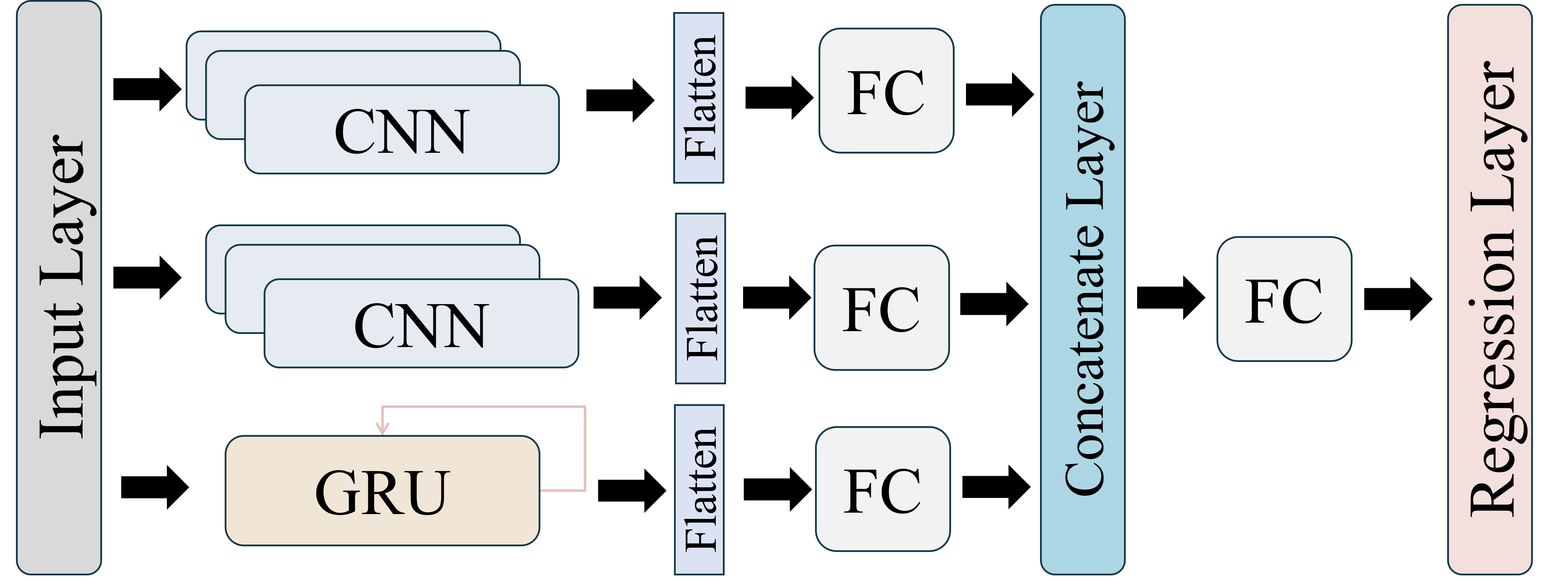}}
    \caption{The Proposed Hybrid CNN-GRU Deep Learning Network.}
    \label{model}
    \end{figure}

    % I said before, use clear colours, black for figures, and suitable colours for blocks
    % Don't ever use white again
    
\section{Results and Discussion}
 
In this section, the performance comparison between various blood glucose level estimation methods reveals significant differences in their accuracy and reliability. Fu-Liang Yang \cite{Fu-Liang} 's method (2021) demonstrated a mean absolute error (MAE) of 8.9 mg/dL, a mean absolute percentage error (MAPE) of 8\%, an $R^2$ score of 0.71, and a root mean square error (RMSE) of 12.4 mg/dL. Similarly, Kim, K.-D \cite{Kim} achieved improved performance with an MAE of 7.05 mg/dL, a MAPE of 6.04\%, an $R^2$ score of 0.92, and an RMSE of 10.94 mg/dL. These metrics indicate a better estimation capability compared to Yang's method, particularly in the $R^2$ score, which reflects the proportion of variance captured by the model.

In contrast, our proposed method (2024) significantly outperforms both aforementioned approaches. With an MAE of 2.96 mg/dL, a MAPE of 2.40\%, an $R^2$ score of 0.97, and an RMSE of 3.94 mg/dL, it demonstrates superior accuracy and precision. The low MAE and MAPE values suggest that our method provides more reliable and closer estimates to the actual blood glucose levels. Additionally, the high $R^2$ score indicates that our model explains a greater proportion of the variance in the blood glucose levels, and the minimal RMSE value highlights its reduced prediction errors. This comparison underscores the effectiveness and robustness of our proposed method in estimating blood glucose levels more accurately than the existing methods.

\begin{table}[h]
    \centering
    \caption{the Performance between Blood Glucose Level Estimation Methods.}
    \label{tab:Related Work performance}
    \begin{tabular}{|c|c|c|c|c|}
        \hline
        Author(Year)& MAE   & MAPE &R\textsuperscript{2}  & RMSE \\
         & (mg/dL)& (\%)& Score &  (mg/dL)\\
        \hline
        Fu-Liang Yang (2021)  \cite{Fu-Liang}& 8.9  & 8 & 0.71& 12.4\\
         \hline
        LRCN  (2023) \cite{LRCN}& 4.7  & - & 0.88&  11.46\\
        \hline
        Kim, K.-D(2024) \cite{Kim} & 7.05  & 6.04 & 0.92 &  10.94\\
         \hline
        Our Method (2024)& \textbf{2.96} & \textbf{2.40 }& \textbf{0.97}& \textbf{3.94}\\
         \hline
    \end{tabular} 
\end{table}

 A 10-fold cross-validation approach was applied to assess the proposed model's reliability and performance. The data was split into 10 groups. For each fold, one group was held out for testing, while the remaining nine groups were used for training and validation. This process was repeated 10 times, ensuring that all data points were used for testing once. Table IV shows the 10-fold cross-validation metrics results. The model's performance on testing data was evaluated using two metrics, i.e., MAE and RMSE. The MAE ranges from 0.76 mg/dL, which is the lowest and shown in fold 5, to 3.75 mg/dL, which is the highest value and presented in fold 1. Meanwhile, the RMSE ranges from 1.26 mg/dL to 7.10 mg/dL, which appear in folds 7 and 1 respectively.

% first letter only capital // latex is capitlize with defualt i think IEEE standard
\begin{table}
    \centering
    \caption{The 10-fold cross-validation metrics results}
    \label{tab:10-fold performance}
\begin{tabular}{|c|c|c|}
\hline
\textbf{Fold No.} & \textbf{MAE (mg/dL)} & \textbf{RMSE (mg/dL)} \\ \hline
1 & 3.759 & 7.107   \\ \hline
2 & 1.477 & 1.7395 \\ \hline
3 & 3.823 & 4.5719  \\ \hline
4 & 1.639 & 2.5983  \\ \hline
5 & \textbf{0.767} & 1.3486 \\ \hline
6 & 1.963 & 2.9789 \\ \hline
7 & 1.041 & \textbf{1.2684}  \\ \hline
8 & 1.029 & 1.8527 \\ \hline
9 & 0.980 & 1.3221 \\ \hline
10 & 1.213 & 1.4846 \\ \hline
\end{tabular}
\end{table}

\begin{figure}[tb]
\centerline{\includegraphics[width=0.3\textwidth]{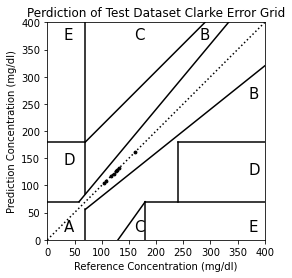}}
\caption{Clarke Error grid on test samples.}
\label{Clarke Error grid}
\end{figure}

Moreover, The Clarke Error Grid (CEG) was used to evaluate the clinical accuracy of the proposed system as shown in Figure \ref{Clarke Error grid}. On the x-axis, the reference BGL is plotted, while on the y-axis the predicted BGL is plotted. The CEG consists of five zones labeled from A to E. Zone A corresponds to Clinically reliable  measurements, while zone E indicates misleading measurements and was implemented in Python with the help of \cite{clarke_Github}.

100 \% of the predictions of test samples analyzed using CEG fall within zone A. This means that the predicted values are within a range of $\pm 20\%$ of the ground truth measures, or both the predicted and reference values are less than 70 mg/dL. This result indicates the clinical safety of the proposed system \cite{clarke}.

% one more result?? // don't understand |there's a test called Clarke Error grid to accept gluceose devices|

\section{Conclusion}
A painless and cheaper non-invasive blood glucose level monitoring system was proposed in this paper overcoming the problems of the conventional fingertip pricking 
approach.

In this study, a hybrid CNN-GRU deep learning network is proposed, the proposed model combines CNN for revealing hidden features relevant to estimating BGL and GRU with memory capabilities to capture temporal relationships between these features.

%The experimental results showed that the suggested model outperformed the state-of-the-art methods, showing promising results achieving an MAE of 2.96 mg/dL, a MAPE of 2.40\%, an $R^2$ score of 0.97, and an RMSE of 3.94 mg/dL

The proposed method is an initiative for applying the combination of convolution and recurrent neural networks to monitor blood glucose levels. The suggested model outperforms the state-of-the-art methods, showing promising results concerning MAE, MAPE, $R^2$ score, and RMSE metrics, 2.96 mg/dL, 2.40\%, 0.97, and 3.94 mg/dL respectively according to the provided dataset.

However, due to the range of the used dataset, i.e., in the range of 88-187 mg/dL, significant modifications and consideration should be made to employ it in care units and hospitals. Consequently, in future extensions, for a reliable system, a new dataset will be constructed with a wider range of readings and more samples to represent the different cases of BGL.
% first time BGL, but looks werid??? idk // it's used alot here n this copy 


\begin{thebibliography}{00}
\bibitem{WHO}Diabetes. (2019, May 13). https://who.int/health-topics/diabetes(Accessed: 26 May 2024).
\bibitem{ppg}Jindal, G. D., et al. ''Non-invasive assessment of blood glucose by photoplethysmography.'' IETE Journal of Research 54.3 (2008): 217-222. https://doi.org/10.1080/03772063.2008.10876202
\bibitem{Fu-Liang} Chu, Justin, Wen-Tse Yang, Wei-Ru Lu, Yao-Ting Chang, Tung-Han Hsieh, and Fu-Liang Yang. 2021. ''90\% Accuracy for Photoplethysmography-Based Non-Invasive Blood Glucose Prediction by Deep Learning with Cohort Arrangement and Quarterly Measured HbA1c'' Sensors 21, no. 23: 7815. https://doi.org/10.3390/s21237815
\bibitem{Kim}Satter, Shama, Mrinmoy Sarker Turja, Tae-Ho Kwon, and Ki-Doo Kim.  ''EMD-Based Noninvasive Blood Glucose Estimation from PPG Signals Using Machine Learning Algorithms'' 2024 Applied Sciences 14, no. 4: 1406. https://doi.org/10.3390/app14041406
\bibitem{LRCN}C. -Y. Liao and W. -C. Fang, ''LRCN-based Noninvasive Blood Glucose Level Estimation,'' 2023 IEEE International Symposium on Circuits and Systems (ISCAS), Monterey, CA, USA, 2023, pp. 1-5, doi: 10.1109/ISCAS46773.2023.10182141.
\bibitem{VGG}Simonyan, Karen, and Andrew Zisserman. ''Very deep convolutional networks for large-scale image recognition.'' 2014 arXiv preprint arXiv:1409.1556.

\bibitem{Lenet}C. Szegedy et al., ''Going deeper with convolutions,'' 2015 IEEE Conference on Computer Vision and Pattern Recognition (CVPR), Boston, MA, USA, 2015, pp. 1-9, doi: 10.1109/CVPR.2015.7298594. 
\bibitem{lstm_vs_gru}Chung, Junyoung, et al. ''Empirical evaluation of gated recurrent neural networks on sequence modeling.'' arXiv preprint arXiv:1412.3555 (2014).
\bibitem{dataset} Kermani Ali, Esmaeili, Hossein. , “The dataset of photoplethysmography signals collected from a pulse sensor to measure blood glucose level” 2023 Mendeley Data, V2, doi: 10.17632/37pm7jk7jn.2
\bibitem{clarke} W.L. Clarke, D. Cox, L.A. Gonder-Frederick, W. Carter, S.L. Pohl, Evaluating 
clinical accuracy of systems for self-monitoring of blood glucose, Diabetes Care 10 
(September (5)) (1987) 622–628, https://doi.org/10.2337/diacare.10.5.622.
\bibitem{clarke_Github} suetAndTie. “GitHub - suetAndTie/ClarkeErrorGrid: This Has the Function for the Clarke Error Grid.” GitHub, n.d. https://github.com/suetAndTie/ClarkeErrorGrid. (Accessed: 2 July 2024)
\end{thebibliography}
\end{document}